\documentclass{aa}

\usepackage{graphicx}

\begin{document}

\newcommand{\MU}{$\mu$m~}
\newcommand{\Tef}{T$_{\rm eff}$~}
\newcommand{\Vt}{$V_{t}$~}
\newcommand{\EBV}{E$_{\rm B-V}$~}
\newcommand{\CC}{$^{12}$C/$^{13}$C}

   \title{
Models of infrared spectra of Sakurai's Object (V4334 Sgr) in 1997
   \thanks{Based  on  observations obtained at the United Kingdom Infrared
Telescope (UKIRT), which is operated by the Joint Astronomy Centre on
behalf of the U. K. Particle Physics and Astronomy Research Council.}}

\author{Yakiv V. Pavlenko\inst{1,2} \and T. R. Geballe \inst{3}}

\offprints{Ya.V. Pavlenko}
\mail{yp@mao.kiev.ua}

\institute{Main Astronomical Observatory of Academy of Sciences of
Ukraine, Golosiiv woods, Kyiv-127, 03680 Ukraine, e-mail:
yp@mao.kiev.ua
\and Isaac Newton Institute of Chile, Kiev Branch
\and Gemini Observatory, 670 N. A'ohoku Place, Hilo, HI 96720 USA,
e-mail: tgeballe@gemini.edu}

\date{Received ; accepted }

\authorrunning{Yakiv V. Pavlenko and T. R. Geballe}
\titlerunning{IR spectrum of V4334 Sgr in 1997}

\abstract{Theoretical spectral energy distributions computed for a grid of
hydrogen-deficient and carbon-rich model atmospheres have been compared
with the observed infrared (1--2.5~$\mu$m) spectra of V4334 Sgr
(Sakurai's Object) on 1997 April 21 and July 13. The comparison yields an
effective temperature of \Tef = 5500 $\pm$ 200~K for the
April date and \Tef = 5250 $\pm$ 200 K for July. The observed spectra are
well fitted by Asplund et al. (1999) abundances, except that the carbon
abundance is higher by 0.3 dex. Hot dust produces significant excess
continuum at the long wavelength ends of the 1997 spectra.
\keywords{Stars: individual: V4334 Sgr (Sakurai's Object) -- Stars:
AGB and post-AGB evolution -- Stars: model atmospheres -- Stars:
energy distributions -- Stars: effective temperatures}
}

\maketitle

\section{Introduction}

V4334 Sgr (Sakurai's Object), the ``novalike object in Sagittarius''
discovered by Y.~Sakurai on February 20, 1996 (Nakano et al. 1996) is a
very rare example of extremely fast evolution of a star during a very late
final helium-burning event (Duerbeck \& Benetti 1996). During the first
few months after discovery, Sakurai's Object increased in visual
brightness to V $\sim$ 12$^m$. In 1997 it increased further to V $\sim$
11$^m$. In March 1997 the first evidence of dust formation was seen (
Kimeswenger et al. 1997, Kamath \& Ashok 1999, Kerber et al. 2000). In
early 1998 the optical brightness of Sakurai's Object decreased (
dimming first reported by Liller et al. 1998), but then recovered.
However, during the second half of 1998 an avalanche-like growth of the
dusty envelope occurred, causing a rapid decrease in optical brightness
and the complete {\bf visual} disappearance of the star in 1999. 
At present essentially
only thermal emission by dust can be observed (Geballe et al. 2002). Our
view of the born again star has been completely obscurred by the dust it
has produced.


 Abundance analyses by Asplund et al. (1997, 1999) and Kipper \&
Klochkova (1997) have found peculiarities similar to those of R CrB-like
stars. Asplund et al. (1999) estimate that the logarithmic abundances of
hydrogen, helium and carbon in atmosphere of Sakurai's Object  in
October 1996 were -2.42, -0.02 and -1.62, respectively\footnote{In this
work we will use an abundance scale $\sum N_i$ = 1.}, with hydrogen only
the third most abundant element by number. All of the above studies are
based on optical spectra obtained in 1996. At that time the spectrum of
Sakurai's Object resembled that of an F-supergiant; molecular bands were
absent or very weak. Cooling of the photosphere of Sakurai's Object
resulted in its optical spectrum during 1997 and 1998 resembling those of
C-giants with very strong bands of CN and C$_2$ (Pavlenko,Yakovina \&
Duerbeck 2000). Modeling of some of these optical spectra have allowed
estimates of the changes in \Tef and \EBV to be made during this period of
rapid evolution of the optical spectrum (Pavlenko et al. 2000; Pavlenko \&
Duerbeck 2001).

Modeling of near infrared (1--2.5~$\mu$m) spectra of Sakurai's Object is
of interest for several reasons. In addition to providing comparisons with
results obtained from the optical spectrum and tests of the reliability of
molecular and atomic data, it allows accurate determination of the
effective temperature and sensitive tests for emission by hot dust. Use of
the 1--2.5~$\mu$m region for modeling is especially important after 1996,
when the bulk of the photospheric flux shifted from the optical into this
waveband.

In this paper we present and compare model 1--2.5~$\mu$m spectra with
those of Sakurai's Object obtained during 1997, on UT April 21 and July 13
at the United Kingdom Infrared Telescope (UKIRT).  The observed spectra
together with observational details were presented by Eyres et al. (1998)
and the July spectrum is also shown in Geballe et al. (2002). The
resolutions of these spectra as presented here are 1.4~nm (0.0014~$\mu$m)
at 1.02--1.35~$\mu$m and 2.8~nm (0.0028~$\mu$m) at 1.42--2.52~$\mu$m.  
 Narrow spectral features in the 1.82--1.95~$\mu$m portions of these
spectra are due to incomplete removal of strong telluric lines.

\section{Modeling Procedure}

Grids of plane-parallel model atmospheres in LTE, with no energy
divergence were computed by the SAM12 program (Pavlenko 2002). This
program is a modification of ATLAS12 (Kurucz 1999).  Opacities due to
C~I bound-free absorption, of importance in atmospheres of
hydrogen-deficient, carbon-rich stars over a wide (0.1--8~$\mu$m)
wavelength region (see Pavlenko 1999, 2002; Asplund et al. 2000), were
computed using the OPACITY PROJECT cross sections database (Seaton et al.
1992). The opacity of C$^-$ also was taken into account 
Myerscough \& McDowell (1996); see Pavlenko 1999 for more details).

An opacity sampling approach (Sneden et al. 1976) was used to account for
atomic and molecular line absorption. The source of the atomic line
information was the VALD database (Kupka et al. 1999). Lists of diatomic
molecular lines of $^{12}$CN, $^{13}$CN, $^{12}$C$_2$, $^{13}$C$_2$,
$^{12}$C$^{13}$C, $^{12}$CO, $^{13}$CO, SiH, and MgH were taken from
Kurucz (1993). We adopted Voigt profiles for every absorption line;
damping constants were computed following Unsold (1955). Microturbulent
velocities of  3--6 ~km~s$^{-1}$ were adopted.

Two grids of model atmospheres with different abundances were computed.
One grid used the chemical composition of Sakurai's Object determined by
Asplund et al. (1999)  for October 1996. The other is the same except
that the abundance of carbon is increased by 0.6 dex. This ``carbon-rich''
case is of interest because the carbon abundance of Asplund et al. (1997)
is larger by 0.6 dex than that which was obtained by them later from the
analysis of high resolution spectra (see Asplund et al. 1997, 1999,
2000. This ``carbon problem'' appears to arise more from the
analysis of C~I lines than from C~II or C$_2$ lines (Asplund, private
communication). For both grids the isotopic ratio \CC = 5 was adopted
(Asplund et al. 1997).

To determine molecular densities, a system of equations of chemical
equilibrium was solved for a mixture of $\sim$70 atoms, ions and
molecules, including the most abundant diatomic molecules containing
carbon.  We used the approach developed by Kurucz in ATLAS12 (Kurucz 1999)
in which ratios of densities of atoms and $n_x$, ... ,$ n_z$ and molecules
$n_{xy...z}$ follow the equations:

\begin{eqnarray}
n_x*...*n_z/n_{x...z}= 
exp(-D_{0}/T_{ev}+b- \nonumber\\
c* (T+d*(T-e*(T+f*T)))+ \nonumber\\ 
3/2*(m-k-1)*ln T),
\label{eq3}
\end{eqnarray}
where $D_0$ and $T_{ev}$ are the dissociation potential and the
temperature in eV (for neutral diatomic molecules m=1, k=0). The adopted
molecular constants for the three most important molecules are given in
Table~1.

\begin{table*}
 \centering
 \begin{minipage}{140mm}

\caption{Dissociation equilibrium constants of CO, C$_2$ and 
CN}
\begin{tabular}{ccccccc}
\hline\hline
\noalign{\smallskip}
Molecule & $D_{0}$ & $b$&$c*$1.E2&$d*$1.E6&$e*$1.E10&$f*$1.E15 \\
\hline
\noalign{\smallskip}
      &        &       &       &       &        &       \\
C$_2$ &  6.116 & 48.75 & .2192 & .4149 & .4121  &1.550 \\
CN    & 7.700  & 47.45 & .1332 & .1989 & .1778  & .6323 \\
CO    & 11.105 & 49.45 & .1651 & .3103 & .3100  &1.168  \\
      &        &       &       &       &        &      \\
\hline
\hline
\noalign{\smallskip}
\end{tabular}
\end{minipage}
\end{table*}

Synthetic spectra\footnote{In fact these are spectral energy distributions
(SEDs).} were computed using the WITA6 program (Pavlenko 1999) for the
same grid of opacities, abundances, isotopic ratios and microturbulent
velocites that were adopted for the model atmospheres. WITA6 computes
spectra and spectral energy distributions (SEDs) taking into account
``line by line'' absorption by atomic and molecular transitions. For
Sakurai's Object spectra computed with wavelength step 0.05~nm were
convolved with gaussian profiles with full widths at half maximum of 1.4
and 2.8~nm over the appropriate wavelength intervals. Computed and
observed spectra were normalized at 1.7~\MU for comparison; see
Pavlenko, Yakovina \& Duerbeck (2000) for more details.

\section{Results}

\subsection{Principal spectral features}

\begin{figure}
\begin{center}
\includegraphics[width=88mm,height=70mm,angle=00]{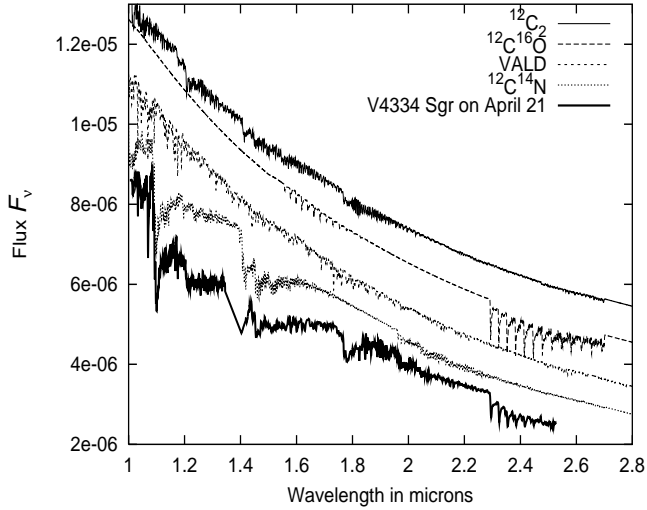}
\end{center}
\caption[]{\label{_01ident_} Model spectra of species that produced the
strongest absoption features in the 1.0-2.5~$\mu$m spectrum of Sakurai's
Object spectra during 1997, computed for \Tef/log g = 5500/0.0 model
atmosphere with Asplund at al. (1999) abundances  for October 1996.
The model spectrum due to atomic species alone, labelled VALD (see
text) is also shown. Spectra are artificially shifted on the y-axis.}
\end{figure}

In Fig. \ref{_01ident_} the principal features formed by the molecular
species C$_{2}$, CO, and CN are displayed in separate spectra. Atomic
features are also shown, as these are also present in Sakurai's Object
(see Eyres et al. 1998; Geballe at al. 2002). As in the optical spectrum
(Pavlenko et al. 2000), absorption of only a few molecular species
accounts for the main features in the IR spectrum. Only the most abundant
isotopic species of each molecule is shown. Of the less abundant isotopic
species, only bands of $^{13}$CO have been detected in the infrared (Eyres
et al. 1998).

\subsection{Dependences on \Tef, log~g, \Vt and log~N(H)}

\begin{figure}
\begin{center}
\includegraphics[width=88mm,height=70mm,angle=00]{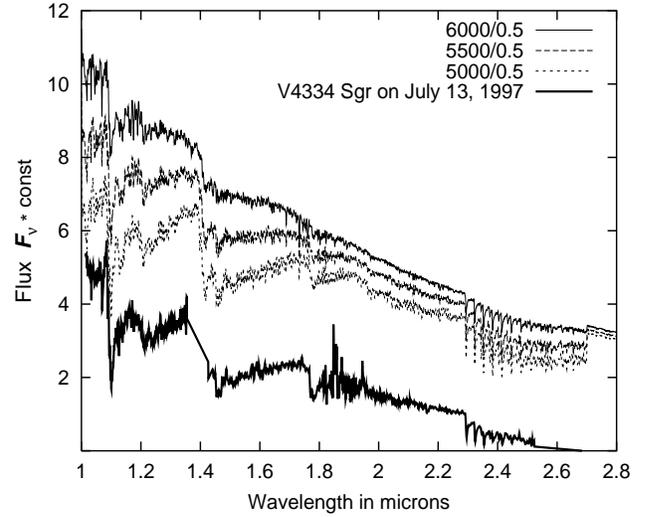}
\end{center}
\caption[]{\label{_02teff_} Dependence of the model IR spectrum on
\Tef. The model spectra use Asplund at al. (1999) abundances
for October 1996. The observed spectrum of Sakurai's Object on
July 13, 1997 is shifted on the y-axis.}
\end{figure}

The model spectra of Sakurai's Object display a strong dependence on \Tef
(Fig. \ref{_02teff_}).\footnote{For $\lambda$~$>$~2.7~$\mu$m model spectra
shown here and in subsequent plots were computed without molecular
absorption; i.e, they are continuum fluxes, which provide information
about the dependence of the continuum fluxes on input parameters.} In
general, the dependence of the IR SED on \Tef is determined mainly by the
variations of the molecular densities with temperature. The band strengths
of CN, CO and C$_{2}$ all increase as \Tef decreases. Changes in the
continuum fluxes are much smaller. Similar effects are seen in model
optical spectra (Pavlenko \& Yakovina 2000). However, there the molecular
bands are numerous, whereas in the infrared only the few strongest
vibration-rotation bands of CN, C$_2$, and CO are prominent.

\begin{figure}
\begin{center}
\includegraphics[width=88mm,height=70mm,angle=00]{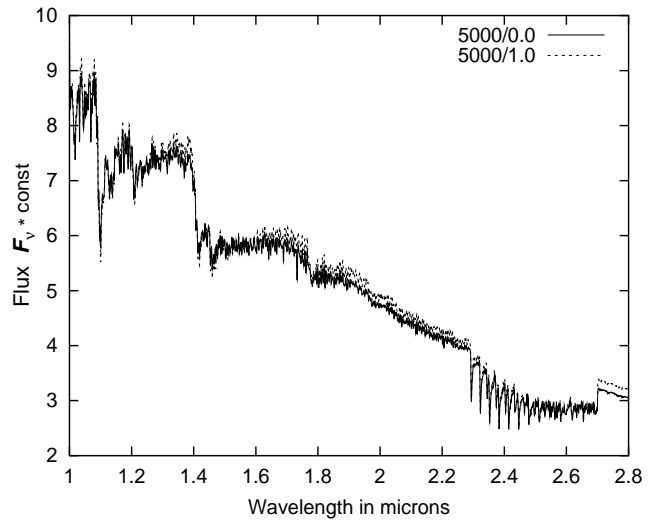}
\end{center}
\caption[]{\label{_03logg_} Dependence of the model spectrum on log g.}
\end{figure}

As can be seen in Fig. \ref{_03logg_}, the dependence of the spectrum on
log g is generally rather weak. However, there are differences in the
responses of different spectral regions.  The strong molecular bands show
rather weak dependence on log g, whereas the fluxes at 1.25-1.35,
1.60-1.75, 1.9-2.2 microns show more noticeable changes.

\begin{figure}
\begin{center}
\includegraphics[width=88mm,height=70mm,angle=00]{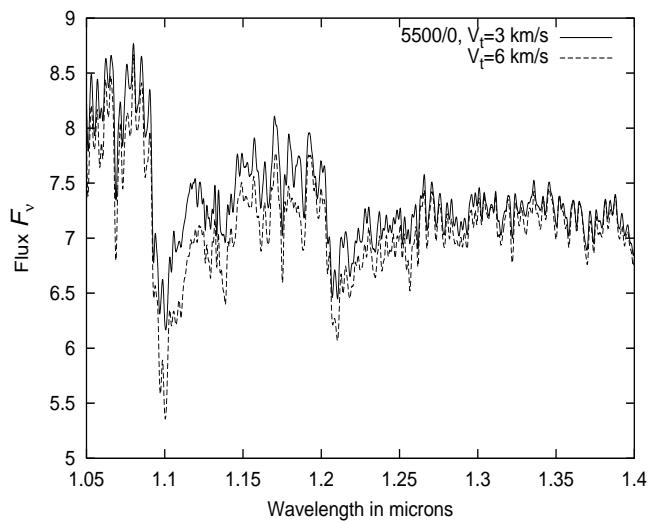}
\end{center}
\caption[]{\label{_vt_} Dependence of the model spectrum on \Vt.}
\end{figure}

Previous abundance analyses of the spectra of Sakurai's Object and
related R CrB stars indicate microturbulent velocities \Vt in the range
5-8 km/s (cf. Asplund et al. 2000). 
The value of \Vt affects the spectral distribution, as is shown
in Fig. \ref{_vt_}. The effect of \Vt on the IR spectra of Sakurai's
Object is larger at the heads of molecular bands than elsewhere, because
the heads are formed by closely packed molecular lines whose overall
absorption is sensitive to \Vt.

\begin{figure}
\begin{center}
\includegraphics[width=88mm,height=70mm,angle=00]{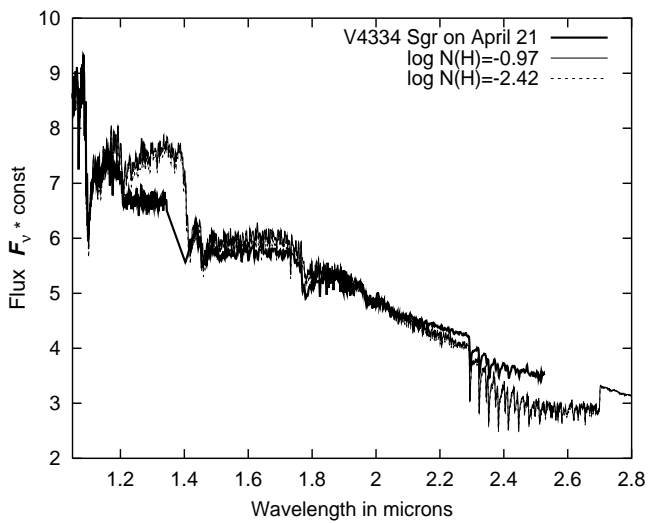}
\end{center}
\caption[]{\label{_04logH_} Dependence of the model spectrum of Sakurai's
Object on log N(H)}
\end{figure}

The main sources of line opacity in the model atmospheres
approximating Sakurai's Object in 1997 are molecular (Pavlenko et al.
2000). Thus it is not surprising that the optical spectra which match
Sakurai's Object respond weakly to changes in the hydrogen abundance.  
This is in contrast to the behavior of models corresponding to the star a
year earlier (Asplund et al. 1997). Similarly, the model IR spectra of
Sakurai's Object for \Tef = 5000--6000~K depend weakly on log N(H) (Fig.
\ref{_04logH_}). The magnitude of the change in the spectrum when log N(H)
is changed from -2.42 (the Asplund et al. 1999 value {\bf for October
1996}) to -0.97 (i.e, a change of 1.5 dex) is comparable (in a qualitative
sense) to lowering log g from 1 to 0 (Fig. \ref{_03logg_}).

\subsection{Changes between 1997 April 21 and July 13}

Fits to the spectra of Sakurai's Object on April 21 and July 13 are shown
in Figs. \ref{_05sak_} and \ref{_06sak_}.  The long wavelength portion of
the H band is of special interest for the ``carbon problem,'' because the
strongest absorption bands of the C$_2$ molecule, the Ballick-Ramsey
bands, occur just longward of 1.768~$\mu$m. In the comparatively hot
atmosphere of Sakurai's Object log N(C) $>$ log N(O) (Asplund et al. 1997,
1999) and the abundance of C$_2$ depends mainly on the elemental abundance
of carbon. Therefore, these bands may provide the most accurate
determination of log~N(C). The fits imply that the carbon abundance is in
the range log N(C) = -1.3 $\pm$0.2. The most likely value is 0.3~dex
higher than that found by Asplund et al. (1999).  The accuracy of the
determination of log N(C) is limited mainly by the quality of the
molecular line list.

The effective temperatures that best fit the 1.0-2.0~$\mu$m spectra in
1997 April and July are 5500~$\pm$~200~K and 5250~$\pm$~200~K,
respectively, indicating that the cooling evidenced by the dramatic
spectral changes seen between 1996 and 1997 (e.g., Geballe et al. 2002)
continued in 1997. Our estimated uncertainties in the above temperatures
are rather large, despite the comparatively good fits to the observed
spectra, because of questions concerning abundances, non-sphericity
effects, and dynamical phenomena, and because of contamination of the
spectra by dust emission (see below).

\subsection{Hot dust}

\begin{figure}
\begin{center}
\includegraphics[width=88mm,height=70mm,angle=00]{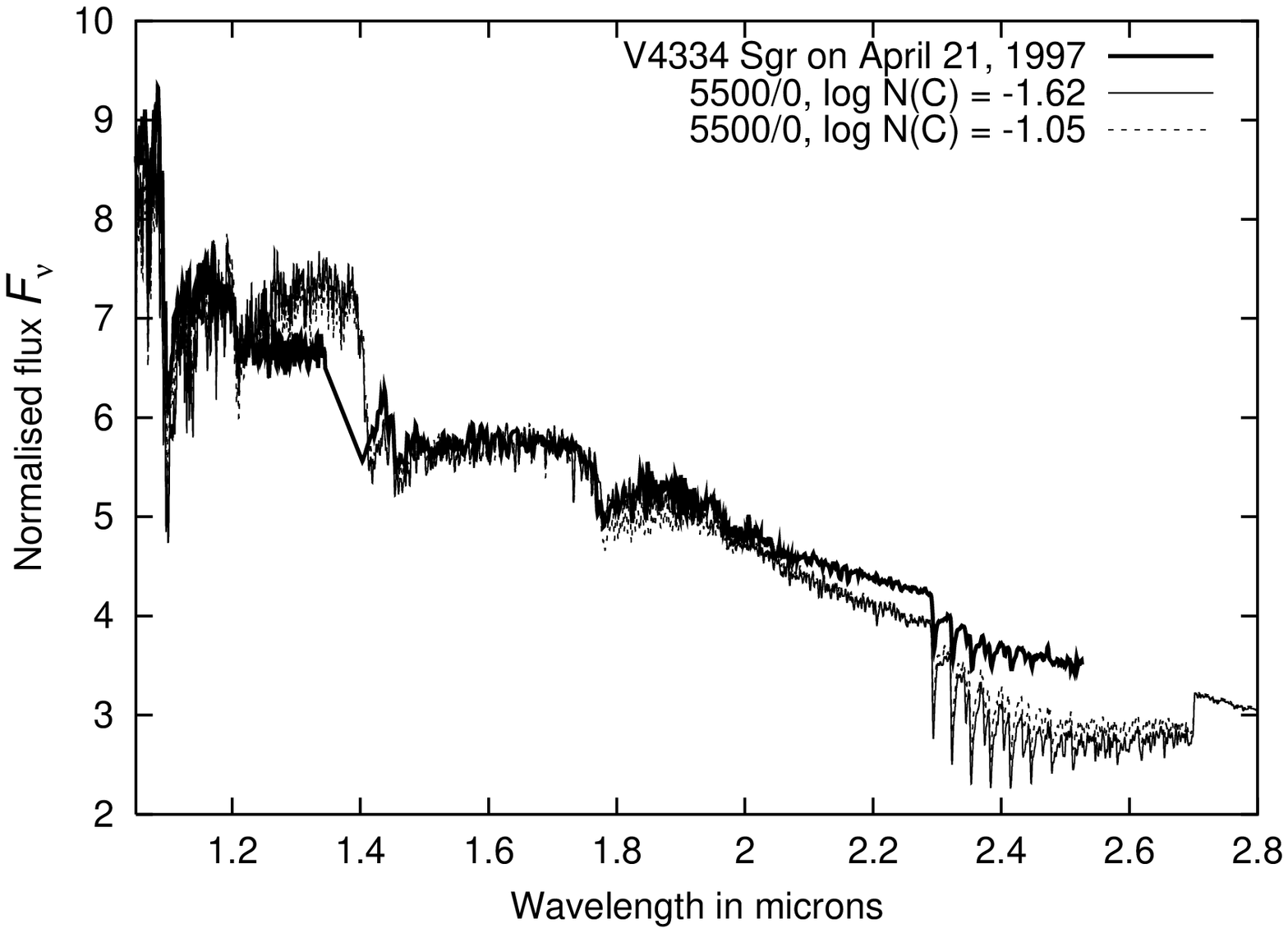}
\includegraphics[width=88mm,height=70mm,angle=00]{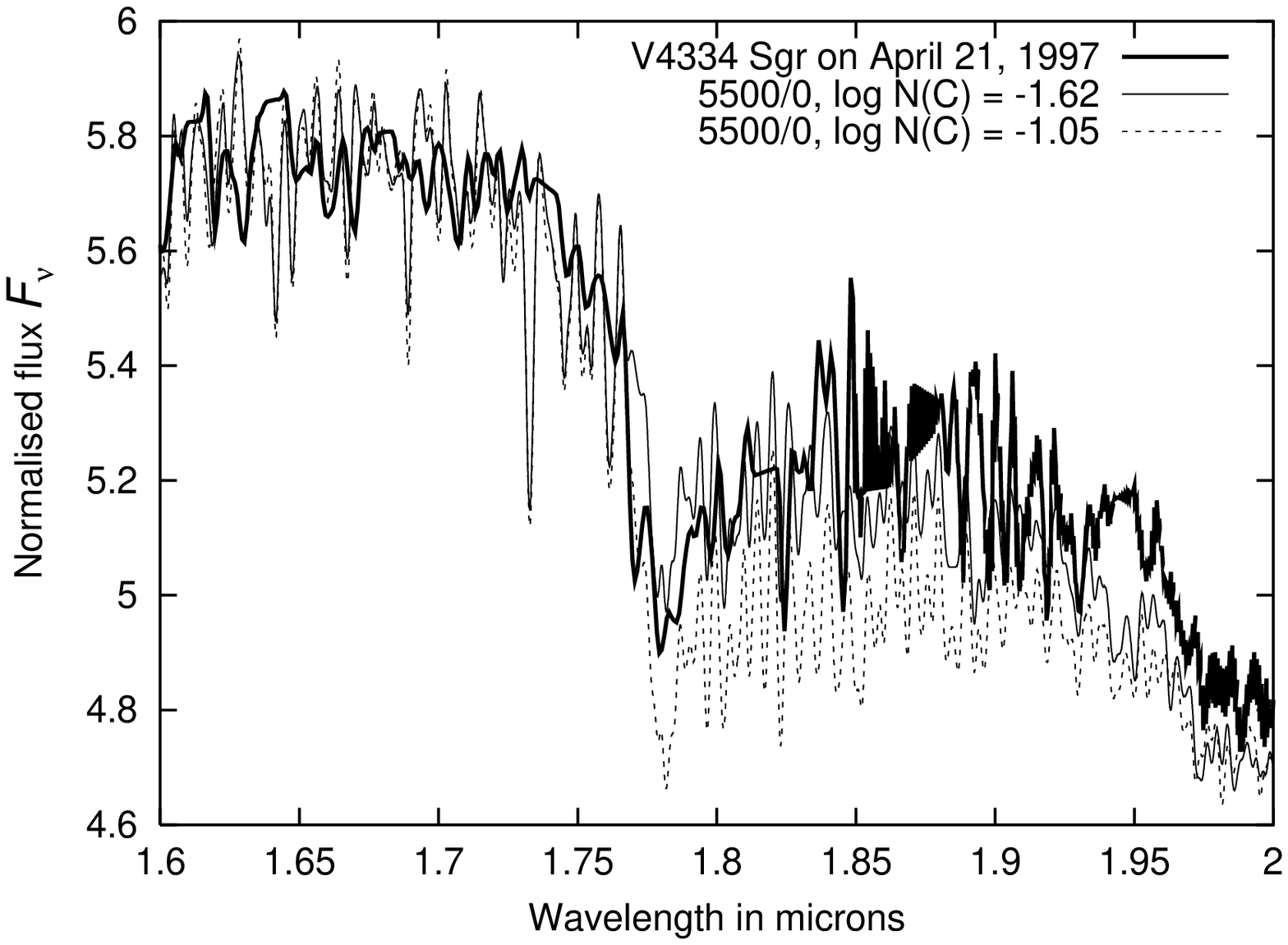}
\end{center}
\caption[]{\label{_05sak_} Top: fits to observed spectrum of of
Sakurai's Object on 1997 April 21. Bottom: details of the fits at
1.6--2.0~$\mu$m; much of the structure at 1.82--1.95~$\mu$m in the
observed spectrum is due to  incomplete removal of telluric
absorption features. Synthetic spectra were computed for a
microturbulent velocity of 6~km/s.}
\end{figure}

\begin{figure}
\begin{center}
\includegraphics[width=88mm,height=70mm,angle=00]{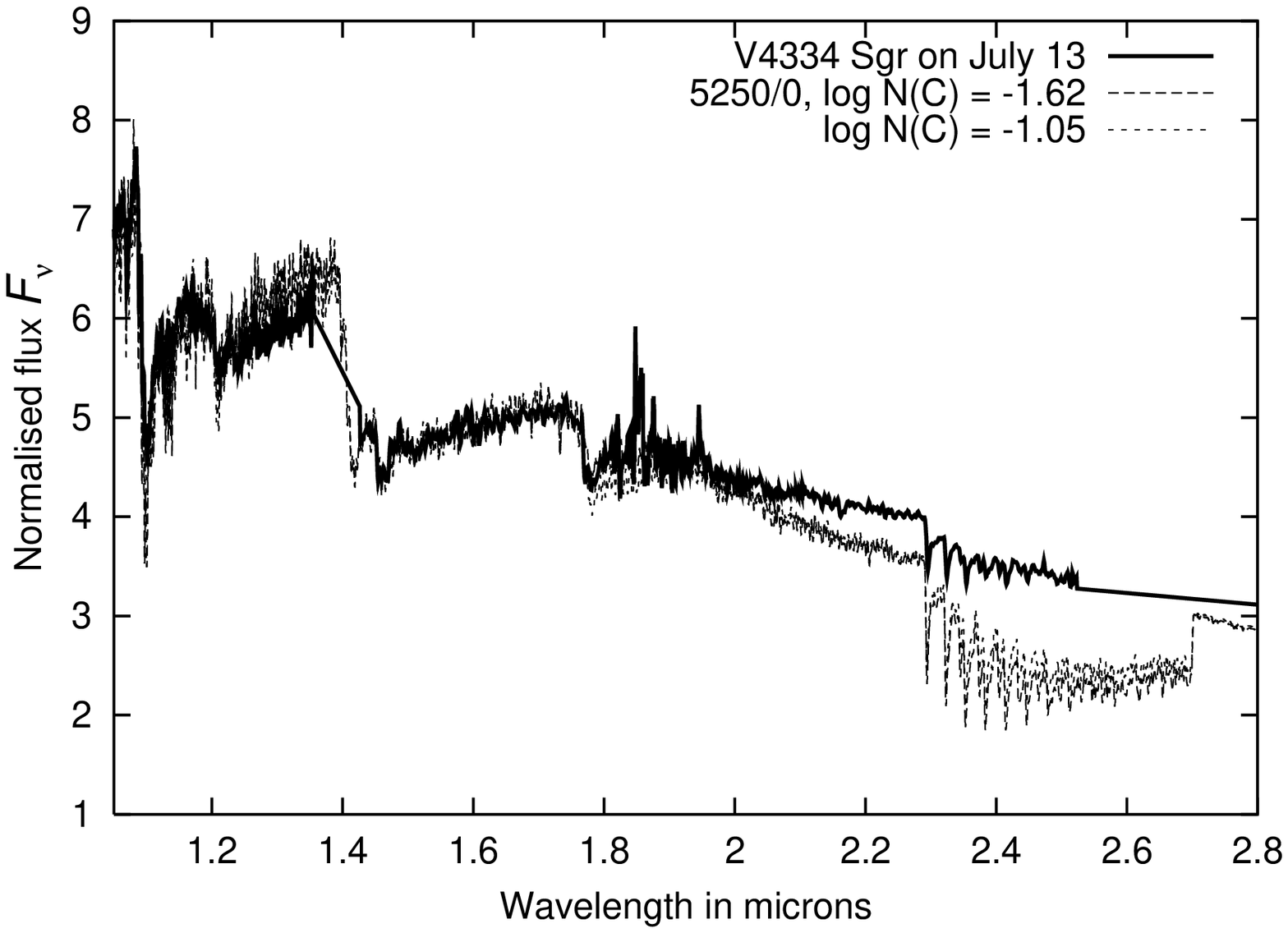}
\includegraphics[width=88mm,height=70mm,angle=00]{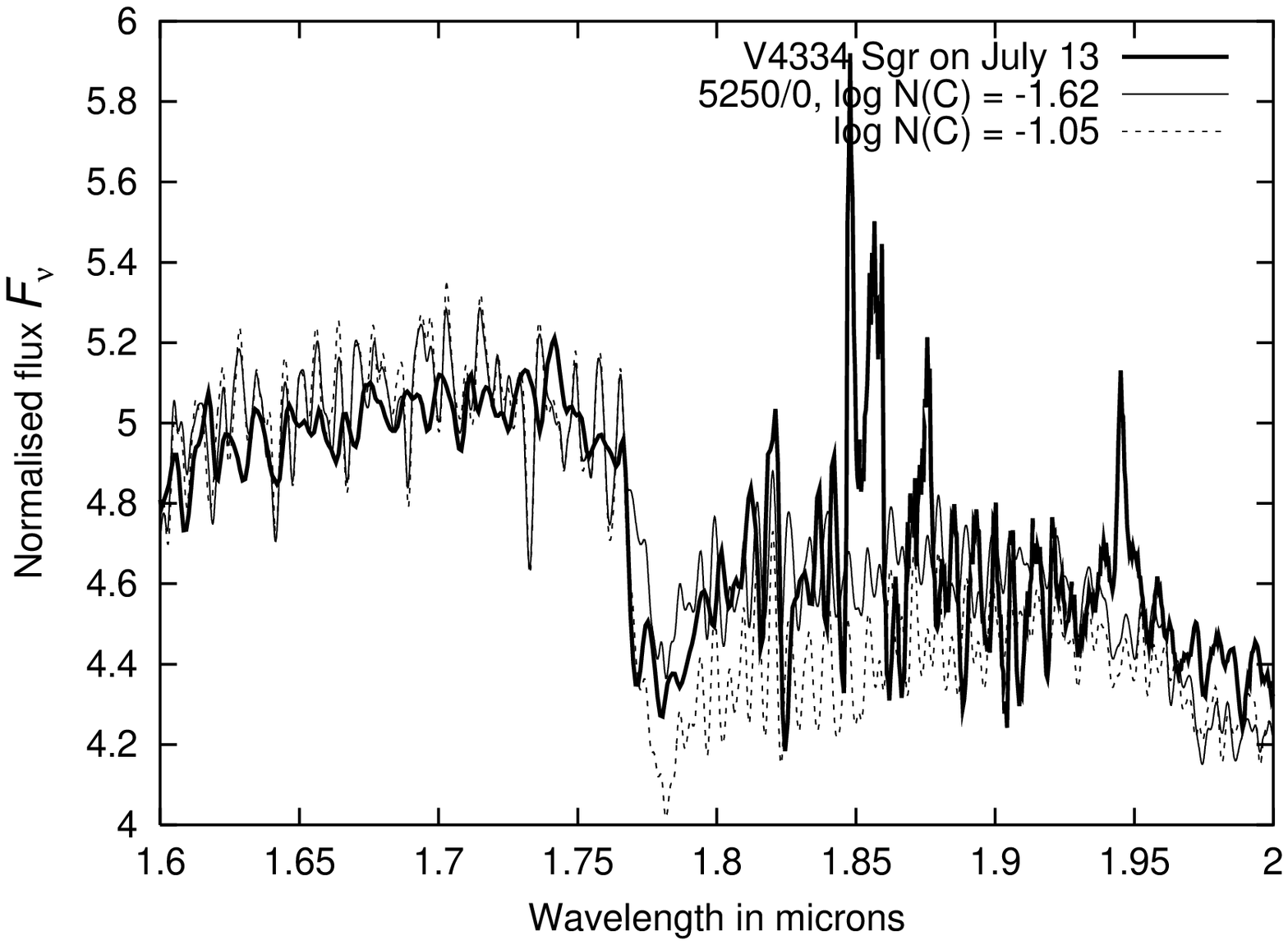}
\end{center}
\caption[]{\label{_06sak_} Top: fits to observed spectrum of of
Sakurai's Object on 1997 July 13. Bottom: details of the fit at
1.6--2.0~$\mu$m. Synthetic spectra were computed for a microturbulent
velocity of 6~km/s.}
\end{figure}

Emission by dust is evident in the 1997 spectra by the mismatch between
the synthetic and observed spectra longward of 2.0~$\mu$m in
Figs. \ref{_05sak_} and \ref{_06sak_}.  The difference
between the observed and synthetic spectra is greater in the July
spectrum, attesting to an increase in the amount of dust.  The thermal
emission from the dusty envelope overlaps the region of first overtone
bands of $^{12}$CO and $^{13}$CO at $\lambda~>~$2.3~$\mu$m. Usually these
bands are used for determination of carbon abundances and isotopic ratios
(cf. Lazarro et al. 1991). The reduced equivalent widths of the CO bands
in July 1997 cannot be reasonably attributed to a large decrease in the
oxygen abundance, because (1) this is unlikely to have occurred in three
months and because the continuum shortward of the CO bands also shows an
excess. We note that in fitting spectra, the most frequent situation is
that the computed spectra have excess flux due to the deficit of known or
hypothetical opacities. To fit the observed spectra, opacities in the
model would need to be {\em decreased} at $\lambda >$ 2~$\mu$m, an
unrealistic possibility.

\section{Discussion}

Our analysis of the 1997 infrared spectra of Sakurai's Object strongly
implies that dust was already present at that time. We note that Duerbeck
(2002) did not find evidence for dust in the optical spectra from 1997. On
the other hand, the fits by Pavlenko \& Duerbeck (2001) to the observed
SEDs at optical wavelengths indicate that \EBV had increased by 0.6 (from
0.7 to 1.3) from April 1997 to August 1998. However, the August 1998 data
were best fit using \Tef~=~5250~$\pm$~200~K, the same value as for July
1997 in this paper. Between  June 1997 and August 1998 there
were some variations in the photospheric radiaton, probably caused by mass
losses events, evolution of the dusty envelope, and dynamical processes in
the photosphere --envelope system (see light curve of Sakurai's Object in
Duerbeck 2002).  Nevertheless, \Tef apparently remained nearly constant
during this period.

In 1997, the year of maximum optical brightness of Sakurai's Object, the
luminosity then was still dominated by optical radiation. At that time
``quasi-periodic fluctuations of increasing cycle length and amplitude
were superimposed on the general brightness evolution'' (Duerbeck 2002).
 In general, the effective temperature during such fluctuations does
not need to follow changes of luminosity. In fact, it can be
anti-correlated, since an increased radius can more than compensate for a
lower \Tef. On the other hand, a change of radius can change the
thermodynamical properties in the radiating region (i.e. photosphere).
That may explain the similarity of \Tef obtained in this paper for July
1997 and that found in August 1998 by Pavlenko \& Duerbeck (2001).  The
decreased optical brightness of Sakurai's Object in 1998 was mainly caused
by development of the dust envelope (Kimeswenger 1999).

One question arises --- were the optical and 2~$\mu$m SED's being affected
by the same dust in 1997-1998? The answer is probably yes. As mentioned
earlier, the effective temperature remained constant during this time and
thus cannot be the cause of the large change in \EBV. This suggests that
the cause of the increase in \EBV was newly formed dust.  The new dust
would be expected to have been close to the star and thus quite hot.
Indeed the full 1--5~$\mu$m spectrum from 1998 (e.g., Geballe et al. 2002)
shows that the excess peaked close to 3~$\mu$m, indicating a mean dust
temperature close to 1,000~K at that time.  The dust must have been hotter
(and closer) in 1997; this is supported by the data from 1997 (Eyres et
al. 1998; Geballe et al. 2002) which show that the continuum flux density
decreased monotonically with wavelength in the observed wavelength range,
1--4~$\mu$m, at that time. Comparison of the 1997 and 1998 SEDs also show
that much less dust was present in 1997. Thus we confirm that the first
appearance of dust occured in 1997 and that the amount of dust increased
through summer 1998, prior to its becoming totally dominant in the latter
part of 1998 and since then.

\begin{acknowledgements}

We thank the staff of the Joint Astronomy Centre for assistance in
obtaining the spectra and the VALD database team for its helpful
assistance. Partial financial support for YVP was
provided by a Small Research Grant from the American Astronomical Society.
TRG's research is supported by the Gemini Observatory, which is operated
by the Association of Universities for Research in Astronomy, Inc., on
behalf of the international Gemini partnership of Argentina, Australia,
Brazil, Canada, Chile, the United Kingdom, and the United States of
America. We thank the referee, M. Asplund for several helpful
suggestions.

\end{acknowledgements}

\end{document}